\documentclass[aps, pra, twocolumn, superscriptaddress, amsmath, amssymb]{revtex4-2}

\makeatletter
\renewcommand{\fnum@figure}{FIG~\thefigure}
\makeatother

\usepackage{graphicx}
\usepackage{dcolumn}
\usepackage{bm}
\usepackage{mathptmx} 
\usepackage{color}
\usepackage[dvipsnames]{xcolor}
\usepackage[hidelinks,colorlinks=true, linkcolor=WildStrawberry, urlcolor=WildStrawberry, citecolor=Emerald, linktocpage=true, breaklinks=true, bookmarksopen=true]{hyperref}
\usepackage{multirow}
\usepackage[USenglish]{babel}

\begin{document}
\title[Body-of-Revolution Finite-Element Model of Plasmon-Enhanced Fluorescence]
{Body-of-Revolution Finite-Element Model of Plasmon-Enhanced Fluorescence} 

\author{Luke C. Ugwuoke}
\email{ lcugwuoke@gmail.com }
\affiliation{
School of Physics and Astronomy, University of Birmingham, Edgbaston, B15 2TT, United Kingdom.}
\author{Farooq Kyeyune}
\affiliation{
Department of Physics, Faculty of Science, Kyambogo University, P.O. Box 1, Kyambogo, Kampala, Uganda.}
\author{Tjaart P. J. Kr\"{u}ger}
\affiliation{ 
	Department of Physics, University of Pretoria, 
	Private Bag X20, Hatfield 0028, South Africa.}
\affiliation{National Institute for Theoretical and Computational Sciences (NITheCS), South Africa.}
\date{\today}
\hyphenpenalty = 1000
\begin{abstract}
Plasmon-enhanced fluorescence (PEF) is one of the most widely investigated optical phenomena in hybrid systems of emitters and plasmonic nanoantennas, with applications ranging from biosensing to single-molecule emission microscopy. However, the computational optimisation of these systems is frequently hindered by the substantial memory and computational costs associated with full three-dimensional (3D) electromagnetic simulations. In this work, we extend finite-element modelling of PEF beyond spherically symmetric geometries using a body-of-revolution finite-element method (BOR-FEM). By exploiting exact or equivalent rotational symmetry, 3D emitter–nanoantenna systems are reduced to computationally efficient 2D formulations while retaining the essential electromagnetic interactions governing excitation and emission. 
We validate the framework against three previously investigated emitter–nanorod systems, including one requiring an equivalent axisymmetric geometric transformation, before applying it to an emitter–core–shell nanorod system. Specifically, we investigate the PEF of the terminal chlorophyll emitter of the major plant light-harvesting complex (LHCII), the most abundant membrane protein on Earth, interacting with a gold core-dielectric shell nanorod with one or two dielectric shells. 
The simulations reveal that the interplay between excitation enhancement, radiative-rate enhancement, and non-radiative Ohmic losses gives rise to four distinct shell-thickness-dependent operating regimes (quenching, enhancement, suppression, and decoupling), with recovery of the intrinsic quantum yield in the decoupling regime. The predicted enhancement factors for experimentally relevant dual-shell nanorods agree well with previously reported measurements. These results establish BOR-FEM as an efficient and versatile framework for modelling PEF in rotationally-symmetric nanoantenna geometries and in non-axisymmetric geometries that admit equivalent axisymmetric representations, providing a practical route for the rational design and optimisation of plasmon-enhanced bio-nanophotonic systems.
\end{abstract}

\maketitle 

\section{Introduction}\label{sec1}
Plasmon-enhanced fluorescence (PEF) arises from the modification of the excitation and emission processes of emitters placed in the near field of metallic or dielectric nanostructures. By tailoring the local density of optical states and restructuring the radiative decay channels, nearby nanoantennas can substantially enhance or suppress an emitter's far-field fluorescence intensity depending on the emitter–nanoantenna geometry, spectral overlap, and separation distance. The phenomenon can be traced back to the pioneering experimental and theoretical works of Drexhage on dipole emitters near dielectric interfaces \cite{Drexhage1970,Drexhage1974}, followed by theoretical investigations of fluorescence modification near nanospheres \cite{Ruppin1982, Ford1984} and nanospheroids \cite{Nitzan1981}. These foundational studies established the framework for what is now commonly referred to as \emph{metal-enhanced fluorescence} (MEF) \cite{Geddes2002,Novot2006,Yaku2026}, or PEF \cite{Polma2007,Halas2009,Khatua2014,Wientjes2014,Bavali2025}. Today, PEF constitutes a central topic in nano-optics and nanophotonics owing to its ability to control excitation rates, spontaneous emission rates through the Purcell effect, and nonradiative decay channels (emission suppression or quenching) \cite{Khatua2014,Wientjes2014,Liaw2012,Niko2016}. 

The ability of plasmonic nanoantennas to tailor light–matter interactions has enabled numerous technological applications. In photovoltaics, PEF has been used to modify the absorption and emission properties of solar cells, thereby improving their power conversion efficiency \cite{Pilla2007,Shen2018,Mandal2022}. Similar concepts have been applied to light-emitting diodes, where plasmonic structures can increase emission efficiency and brightness \cite{Kwon2008,Ma2021}. The strong electromagnetic hotspots associated with plasmonic resonances have also motivated the development of plasmon-enhanced microscopy techniques that offer improved image contrast and spatial resolution \cite{Yelin2003,Hadi2009,Jan2024}. 
In biosensing, plasmonic enhancement has been used to provide enhanced detection sensitivity and signal-to-noise ratios, enabling applications ranging from point-of-care diagnostics and single-molecule detection \cite{Toma2014,Mino2020,Gupta2024,Xu2025} to real-time monitoring of binding kinetics \cite{Okholm2024,Boom2024}.

To understand and optimise these systems, extensive theoretical and computational studies have been devoted to calculating PEF in a wide variety of emitter–nanoantenna geometries. Representative examples include emitters near nanospheres \cite{Del2024,Guza2012} and nanorods \cite{Halas2009, Khatua2014,Moha2008,Liaw2012,Ugwuoke2021,Liu2015,Zhang2018,Nepal2013,Li2010}, emitters in dimer nanoantenna gaps \cite{Moha2008,Lu2022,Kinka2009,Zhang2015,Regmi2016,Musk2007}, emitters inside \cite{Ming2009,Singh2017} and outside \cite{Halas2009,Grady2008,Ugwuoke2020,Arruda2017,Niko2016} nanoshells, and emitters near more complex nanostructures \cite{Martins2012,Chen2012,Peng2014,Zaki2017}. PEF has also been investigated in biohybrid systems involving photosynthetic light-harvesting complexes coupled to plasmonic nanoantennas \cite{Worm2008,Wientjes2014,Bujak2014,Kyeyune2019}. Quantitative modelling of these systems generally requires solving Maxwell's equations in three dimensions (3D) using numerical approaches such as the finite-difference time-domain (FDTD) method \cite{Wientjes2014,Liaw2012}, the boundary element method (BEM) \cite{Vesseur2010,Okholm2024,Boom2024}, the multiple multipole method \cite{Liaw2012,Liaw2011}, discrete dipole approximation (DDA) \cite{Khatua2014}, and quasinormal modes analysis \cite{Yan2018,Wu2023}. 

For emitter–nanorod systems with rotational symmetry, the computational cost of full 3D simulations can be significantly reduced by exploiting the system's symmetry. An important example is the body-of-revolution (BOR) FDTD formulation introduced in Ref. \cite{Moha2008}, which transforms the 3D emitter--nanorod problem into an equivalent 2D representation for an emitter whose dipole moment is aligned with the nanorod's long axis, i.e., the model is valid as long as the dipole emitter is not arbitrarily oriented and the nanoantenna does not lack rotational symmetry. 
This reduction substantially decreases the computational requirements while retaining the essential physics of the emitter–nanoantenna interaction. Nevertheless, most PEF studies continue to rely on full 3D simulations, particularly when evaluating excitation-rate enhancement, Purcell enhancement, antenna efficiency, and fluorescence enhancement over large parameter spaces. As a result, optimisation studies involving variations in nanoantenna geometry, shell thickness, emitter position, or emission wavelength can remain computationally demanding despite the presence of exploitable symmetry.

A further challenge arises in biohybrid PEF systems involving complex multichromophoric emitters such as photosynthetic light-harvesting complexes. In conventional simulations, the emission stage is typically represented by an electric dipole source whose dipole moment must be specified a priori 
~\cite{Khatua2014,Wientjes2014,Moha2008}. For multichromophoric systems, however, assigning a unique dipole moment can be nontrivial because different pigments may contribute to excitation and emission processes \cite{Wientjes2014,Kyeyune2019}. Consequently, care must be taken when translating molecular-scale photophysics into classical electromagnetic simulations.

To address these challenges, we extend the BOR approach to the finite-element method (FEM) for modelling PEF. Specifically, we employ the 2D axisymmetric solver within the frequency-domain Wave Optics module of COMSOL Multiphysics$^{\text{\textregistered}}$~\cite{Comsol2025}. We use the abbreviation BOR-FEM to keep the method nomenclature consistent with other works involving studies such as dielectrometry \cite{Feng2023}, focusing optics \cite{Genti2019}, and photon collection optics \cite{Ander2018,Komi2021}. The resulting framework enables the calculation of excitation enhancement, Purcell enhancement, quantum-yield modification, and fluorescence enhancement using a 2D axisymmetric formulation while preserving the 3D rotationally symmetric characteristics of the system, and without requiring the electric dipole moment of the emitter to be specified explicitly as a simulation input parameter. Because COMSOL Multiphysics is amongst the most widely used finite-element platforms in nanophotonics, the BOR-FEM implementation provides an accessible and computationally efficient route for modelling PEF in rotationally symmetric nanoantenna geometries.

To demonstrate the capabilities of BOR-FEM, we investigate PEF in a hybrid system comprising the terminal chlorophyll (Chl) emitters of the major plant light-harvesting complex (LHCII) and gold core–shell nanorods coated with either single or double dielectric shells. Before applying the framework to this biohybrid system, the method is validated against three experimentally and numerically characterised emitter--nanorod benchmark systems reported in Refs.~\cite{Khatua2014,Boom2024,Wientjes2014}, enabling direct comparison with published DDA, BEM, and FDTD simulations. LHCII provides an attractive demonstration system because of its central role in photosynthetic light harvesting, its status as the most abundant fluorescing protein complex on Earth, its use in biohybrid light-harvesting devices, and the broader interest in controlling excitonic systems using plasmonic nanoantennas. Although the LHCII--nanoantenna system considered here has previously been investigated experimentally~\cite{Kyeyune2019}, to our knowledge it has not been studied using full PEF simulations. By comparing the simulated enhancement factors with experimentally reported values, we evaluate the predictive capability of BOR-FEM while demonstrating a computational framework that is applicable not only to rotationally symmetric emitter--nanoantenna systems but, more generally, to problems that can be represented through an equivalent axisymmetric transformation.

\section{Model}\label{sec2}
\subsection{Model equations in plasmon-enhanced fluorescence}\label{subsec2.1}
\begin{figure}
	\centering 
	\includegraphics[width = .35\textwidth]{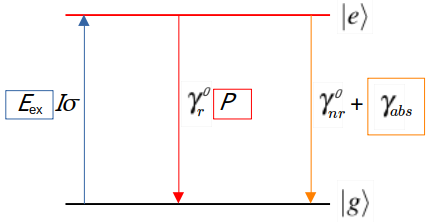}
	\caption{Transition rates of a two-level (with ground state $|g\rangle$ and excited state $|e\rangle$) dipole emitter modified by the plasmon-induced parameters $E_{ex}, P,$ and $\gamma_{abs}$. Adapted from Ref. \cite{Khatua2014}. 
	}\label{f1}
\end{figure}

We model the PEF dynamics using the semiclassical framework popularised by Novotny and co-workers~\cite{ Novot2006, Novot2007, Novotny2012} and others \cite{Geddes2002, Polma2007,Musk2007, Moha2008,Kinka2009,Khatua2014}. In this framework, PEF is treated as a two-stage process comprising excitation and emission. During excitation, the local field generated by the nanoantenna modifies the excitation rate of the emitter. In the subsequent emission phase, the nanoantenna alters the radiative and non-radiative decay channels of the excited emitter, thereby modifying its fluorescence quantum yield and emission rate. The electromagnetic power radiated into the far field and the power lost via Ohmic dissipation within the lossy metal are designated as $P_{rad}$ and $P_{dis}$, respectively. 

The modified transition rates of the dipole emitter are shown in Fig.~\ref{f1}. 
In the excitation stage, the excitation rate of the emitter is given by $E_{ex}I\sigma$, where $E_{ex}$ is the excitation-rate enhancement due to the interaction of the emitter with the nanoantenna, $I$ is the incident photon flux, and $\sigma$ is the absorption cross-section of the emitter. 
In the emission stage, the radiative and non-radiative decay rates of the emitter are modified as $\gamma^0_r P$ and $\gamma^0_{nr} + \gamma_{abs}$, respectively \cite{Novot2007,Khatua2014}, where $\gamma^0_r$ and $\gamma^0_{nr}$ are the radiative and non-radiative decay rates of the emitter in the absence of the nanoantenna, $P$ is the radiative decay rate enhancement, also referred to as the \emph{Purcell factor}, and $\gamma_{abs}$ is the rate of energy absorbed by the nanoantenna in the form of Ohmic heating due to the power dissipated by the emitter. 

The steady-state emission rates of the emitter in the absence ($\gamma^0_{em} $) and presence ($\gamma_{em}$) of the nanoantenna are obtained, respectively, as \cite{Khatua2014} 
\begin{subequations}
	\begin{align}
		\gamma^0_{em} &  = I\sigma\rho^0_g Y_0,  \label{e1}\\
		\gamma_{em} &  = E_{ex}I\sigma\rho_g Y, \label{e2}
	\end{align}
\end{subequations}
where $\rho^0_g$ and $\rho_g$ are the ground-state populations in the absence and presence of the nanoantenna, respectively, and $Y_0 = \gamma^0_r/(\gamma^0_r+\gamma^0_{nr})$ and $Y = \gamma^0_rP/(\gamma^0_rP + \gamma^0_{nr} + \gamma_{abs})$ are the quantum yields of the emitter in the absence and presence of the nanoantenna, respectively. The fluorescence enhancement factor, $F = \gamma_{em}/\gamma^0_{em}$, is obtained under weak-excitation conditions ($\rho^0_g \approx 1, \rho_g \approx 1$, when saturation effects are ignored) as
\begin{equation}
	F \approx E_{ex}(\lambda_{ex}) Y(\lambda_{em})/Y_0, 
\end{equation}
where $\lambda_{ex}$ is the excitation wavelength of the emitter-nanoparticle system and $\lambda_{em}$ is the emission wavelength of the emitter. 
The modified quantum yield $Y$, also referred to as the \emph{total quantum efficiency} of the emitter \cite{Wientjes2014}, is sometimes rewritten in terms of $P$, the antenna efficiency $\eta = P_{rad}/(P_{rad} + P_{abs})$, and the intrinsic quantum yield $Y_0$ as \cite{Moha2008,Novot2007}
\begin{equation}\label{e3}
	Y = Y_{0}\left[ \frac{1-Y_{0}}{P} + \frac{Y_{0}}{\eta} \right]^{-1},
\end{equation}
and the normalised powers and rates are equal, i.e., $P = P_{rad}/P_{rad}^0 = \gamma_r/\gamma_r^0$ and $P_{abs} = P_{dis}$, with $P_{abs}/P_{rad}^0 = \gamma_{abs}/\gamma_r^0$, $P_{rad}^0$ being the power radiated by the emitter in the absence of the nanoantenna \cite{Moha2008,Novot2007}. 

\subsection{Body-of-revolution finite-element method (BOR-FEM)}
\begin{figure}
	\centering 
	\includegraphics[width = .30\textwidth]{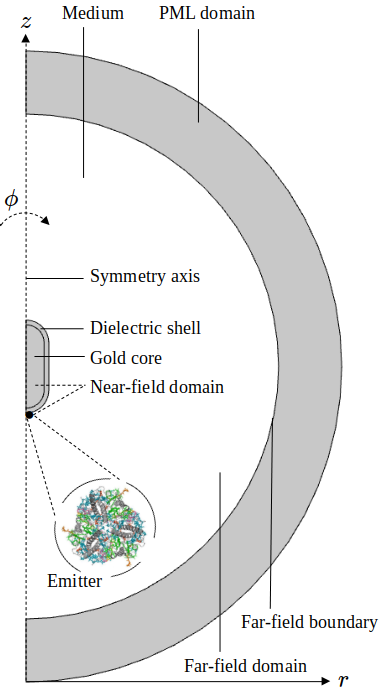}
	\caption{2D model geometry of the emitter-nanorod system in the BOR-FEM simulation. The emitter is assumed to have a $z$-oriented dipole moment. The system is rotationally-symmetric in the $r, \phi, z$ cylindrical coordinate basis. 
	}\label{f2}
\end{figure}

The model system considered in this work consists of the major plant light-harvesting complex, LHCII, interacting with a gold core–dielectric-shell nanorod, with the 2D model geometry shown in Fig.~\ref{f2}. The terminal-emitter Chl cluster of LHCII contributes approximately 97\% to the lowest exciton state \cite{Raman2015}, constituting the primary fluorescence-emitting site of this pigment-protein complex. Consequently, LHCII can be approximated as a single effective dipole emitter for the purposes of the present electromagnetic analysis. The inset in Fig.~\ref{f2} shows the stromal side of LHCII \cite{Liu2004}.  The dielectric shell surrounding the gold core consists of either polystyrene sulfonate (PSS) or cetyltrimethylammonium bromide (CTAB), consistent with previous experimentally studied nanoantenna geometries \cite{Kyeyune2019}. 

The excitation enhancement was determined from the local electric field generated by the nanorod under external illumination. Owing to the rotational symmetry of the system, the incident field was represented by the azimuthally symmetric mode ($m=0$), corresponding to a $z$-polarised cylindrical wave. The scattering response of the core-shell nanorod was first calculated, after which the electric field at the emitter position was used to determine the excitation-rate enhancement. 

Consider a core-shell nanorod in the presence of a $z$-polarised azimuthally-symmetric incident field approximated by the $m=0$ cylindrical wave with expression $E_0 J_{0}(k_nr)\hat{e}_{z}$, where $E_0$ is the amplitude of the field, $k_n = 2\pi n/\lambda_{ex}$ is the wavenumber of the light in the medium, $J_0(\cdot)$ is the cylindrical Bessel function of azimuthal mode order $m=0$, and $\hat{e}_{z}$ is a unit vector in the $z$ direction. The scattering cross-section of this rod is given by \cite{Grand2019}
\begin{equation}\label{e4}
	\sigma_{sca} =  \frac{1}{2S_{0}}\int_S\Re[ \mathbf{E} \times \mathbf{H}^{*} ] dS, 
\end{equation}
with $S_{0} = E_0^2 n/2Z_{0}$ being the intensity of the incident field, $Z_{0}$ the impedance of free space, $n$ the refractive index of the medium, $\int_S$ a surface integral on the dielectric shell of the revolved geometry (i.e., a line integral at the dielectric shell in 2D), and $\Re[ \mathbf{E} \times \mathbf{H}^{*}]/2$ the time-averaged Poynting vector with electric and magnetic field components $\mathbf{E} = (E_{r}, 0, E_{z})$ and $\mathbf{H}^{*} = (0, H^{*}_{\phi}, 0)$, respectively. The excitation-rate enhancement is evaluated at the emitter position by comparing the local electric-field intensity to that of the incident field. Adapting the surface-integral formulation of Ref. \cite{Grand2019} to the present 2D axisymmetric geometry yields 
\begin{equation}\label{e5}
	E_{ex}(r_0,t) = \frac{\int |\mathbf{E}(r_0,z)|^{2}\delta(z-t)dz}{\int E_{0}^{2}\delta(z-t)dz},
\end{equation}
where $(r_0,t)$ denotes the emitter position on the dielectric shell, $t$ is the thickness of the shell, $r_0$ is the radial component of the emitter's position from the symmetry axis shown in Fig.~\ref{f2}, and $|\mathbf{E}(r_0,z)|$ is the magnitude of the local electric field at point $(r_0,z)$. The Dirac delta function restricts the integration to the emitter location, effectively performing a point evaluation of the field enhancement.  

In the emission stage, the emitter is represented as a point source that drives the electromagnetic response of the emitter--nanorod system. Within the 2D axisymmetric formulation of COMSOL Multiphysics$^{\text{\textregistered}}$~\cite{Comsol2025}, this source is implemented using a magnetic current element, $I_m$, with units in Volts, located near the symmetry axis, because the axisymmetric solver does not permit point sources to be placed directly on the axis. The magnetic current element produces a magnetic dipole moment equivalent to a small electric current-carrying circular loop \cite{Balanis1997, Ander2018, Komi2021}, reproducing the radiation pattern of a vertically oriented electric dipole. The value of $I_m$ is therefore calibrated such that that the radiated power equals the free-space radiated power of the emitter, i.e., $P_{rad}^0 = \gamma^0_r \hbar\omega_{em} = \gamma^0_r hc/\lambda_{em}$, where $\omega_{em}$ the emission angular frequency, $c$ is the speed of light in free space, and $h$ is Planck's constant, allowing us to determine the power radiated by the emitter in the absence of the nanorod. The calibration was performed by adjusting $I_m$ until the simulated radiated power matched the target value of $P_{rad}^{0}$. Since the calibration depends on $r_0$, a different value of $I_m$ is generally required when $r_0$ is changed.

The radiated power was calculated from the time-averaged Poynting vector as \cite{Liaw2012,Del2024}
\begin{equation}\label{e6}
	P_{rad} =  \frac{1}{2}\int_{S}\Re[ \mathbf{E} \times \mathbf{H}^{*}]  dS, 
\end{equation}
where the integration is performed over the far-field boundary of the axisymmetric model and revolved about the symmetry axis to obtain the corresponding 3D power. For the vertically oriented dipole considered here, the BOR formulation corresponds to the azimuthal mode number $m = 0$ \cite{Ander2018,Komi2021}, giving a transverse-magnetic field \cite{Moha2008} with $\mathbf{E} = (E_{r}, 0, E_{z})$ and $\mathbf{H}^{*} = (0, H^{*}_{\phi}, 0)$.  The Ohmic power absorbed within the nanorod was calculated as \cite{Del2024,Grand2019}
\begin{equation}
	P_{abs} = \frac{1}{2}\int_{V}\Re[\mathbf{E} \cdot \mathbf{J}^{*}] dV, 
\end{equation}
where $\mathbf{J}^{*} = (J^{*}_{r}, J^{*}_{\phi}, J^{*}_{z})$ is the complex conjugate of the induced current density and the volume integral $\int_{V}$ is evaluated over the corresponding 2D cross-section of the nanorod in the near-field domain.
 
Open-boundary conditions were implemented using a perfectly-matched layer (PML) at the far-field boundary to prevent artificial back-reflections into the computational domain. Physics-controlled fine meshing was employed throughout both the physical and PML domains, while scattering boundary conditions were applied at the outer PML boundary. The physical domain extended to $3\lambda$, where it was surrounded by a PML of thickness $\lambda$, giving a total computational radius of $4\lambda$. Here, $\lambda=\lambda_{ex}$ during the excitation stage and $\lambda=\lambda_{em}$ during the emission stage.

\section{Results and Discussion}
\begin{figure}
	\centering 
	\includegraphics[width = .23\textwidth]{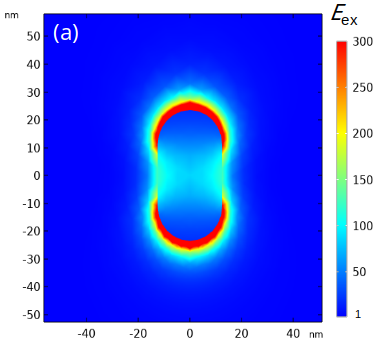}~~~~~~~
	\includegraphics[width = .23\textwidth]{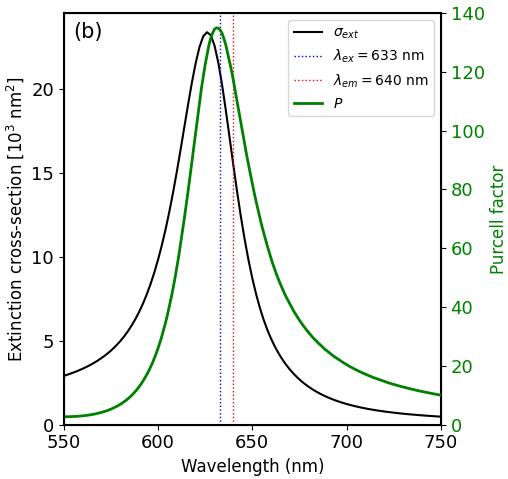} \vspace{0.25cm}\\
	\includegraphics[width = .23\textwidth]{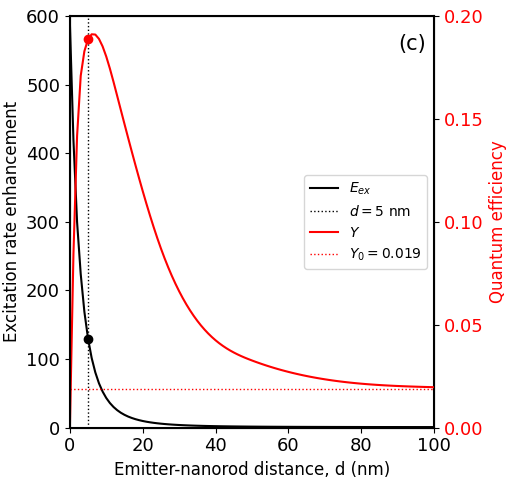}~~~~
	\includegraphics[width = .23\textwidth]{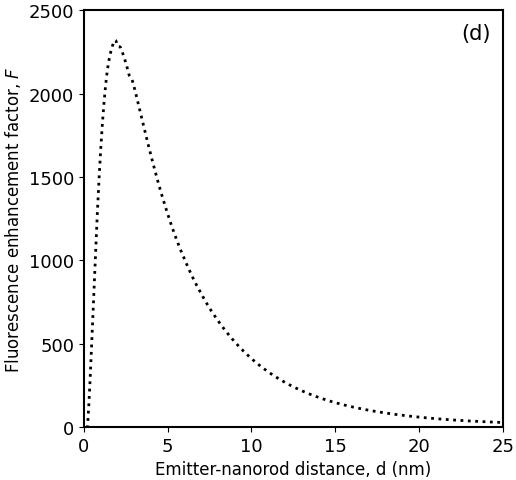}
	\caption{Reproduced results based on BOR-FEM simulations using input data from the CV-AuNR system in Ref. \cite{Khatua2014}. 
		(a) Near-field map of the excitation rate enhancement evaluated at the excitation wavelength $\lambda_{ex} = 633$ nm. (b) Extinction cross-section of the AuNR (black curve) and wavelength dependence of the Purcell factor at $d = 5$ nm (green curve). Excitation and emission wavelengths are shown as blue and red dotted lines, respectively. 
		(c) Dependence of the excitation rate enhancement (black curve) at $\lambda_{ex} = 633$ nm and the modified quantum yield $Y$ (red curve) at the emission wavelength $\lambda_{em} = 640$ nm on the emitter-nanorod separation. The intrinsic quantum yield $Y_0 = 0.019$ is indicated by the dashed horizontal line. 
		(d) Dependence of the fluorescence enhancement factor on the emitter-nanorod distance, evaluated at the excitation and emission wavelengths.
	}\label{fv}
\end{figure} 

\subsection{Validation of the BOR-FEM framework}
Before applying BOR-FEM to the LHCII–nanorod system, we first benchmarked the framework against DDA, BEM, and FDTD calculations of emitter–nanorod systems studied previously. These benchmarks provide a stringent test because they include both excitation- and emission-stage PEF parameters, allowing the complete BOR-FEM workflow to be validated against established computational models.

\subsubsection{Benchmark I: Crystal-violet molecule near a gold nanorod}
 The first system is a single crystal violet (CV) molecule positioned near the tip of a cylindrical gold nanorod (AuNR) immersed in glycerol ($n = 1.47$), which was investigated experimentally and theoretically by Khatua \textit{et al.} using DDA calculations~\cite{Khatua2014}. CV has an intrinsic quantum yield of $Y_0 = 0.019$ and a radiative decay rate of $\gamma_r^0 = 0.019$ ns$^{-1}$ in the absence of the AuNR, while the nanorod dimensions are $L \times W \times W$ with $L=47$ nm and $W=25$ nm, corresponding to an aspect ratio of $L/W = 1.88$. Because this nanorod possesses rotational symmetry, it is well-suited for the axisymmetric BOR-FEM formulation. The refractive index of gold is taken from Ref. \cite{Christy1972}, and the emitter--nanorod separation is denoted by $d$. 

Figure~\ref{fv}(a) compares the excitation-stage near-field obtained using BOR-FEM with that reported by Khatua \textit{et al.} The excitation-rate enhancement map was calculated at the excitation wavelength $\lambda_{ex}=633$~nm for an emitter located 5~nm from the nanorod tip. The 3D field distribution shown in Fig.~\ref{fv}(a) was reconstructed by revolving the 2D BOR-FEM solution about the symmetry axis. The simulated field reproduces both the hotspot localisation and the spatial confinement of the plasmonic near field reported from the DDA calculations, demonstrating that the axisymmetric formulation accurately captures the excitation-stage electromagnetic response.

The spectral response of the nanorod is shown in Fig.~\ref{fv}(b). The extinction cross-section was calculated as $\sigma_{ext} = \sigma_{sca} + \sigma_{abs}$, where $\sigma_{sca}$ is obtained from Eq.~\eqref{e4} and $\sigma_{abs} = P_{abs}/S_0$ is the absorption cross-section determined from the absorbed power during the excitation stage. The resulting extinction spectrum (Fig.~\ref{fv}(b)) yields a longitudinal localised surface plasmon resonance (LLSPR) at $626$ nm, in close agreement with the value of $629$ nm reported in Ref. \cite{Khatua2014}. The extinction spectrum overlaps both the excitation wavelength ($\lambda_{ex}=633$~nm) and the peak fluorescence wavelength ($\lambda_{em}=640$~nm) of the CV molecule. Such dual spectral overlap is well known to maximise PEF because it simultaneously enhances the excitation rate and the radiative emission process~\cite{Novot2007,Wientjes2014}. The wavelength dependence of the simulated Purcell factor, shown by the green curve in Fig.~\ref{fv}(b), is likewise consistent with the DDA calculations and exhibits the expected resonance behaviour.

To determine the emission-stage quantities ($P$ and $Y$), we used $I_m = 1.4385$ V as the calibrated source magnetic current to produce an equivalent radiated power of $P_{rad}^0 = \gamma^0_r hc/\lambda_{em} \approx 5.8973$ pW, corresponding to the free CV molecule placed at $r_{0} = 0.5$ nm from the symmetry axis. This calibration establishes the correct correspondence between the magnetic-current source used in the axisymmetric formulation and the physical electric dipole represented by the CV molecule.

Figure~\ref{fv}(c) shows the dependence of the excitation rate enhancement $E_{ex}$ (black curve) and the modified quantum yield $Y$ (red curve) on the emitter-nanorod distance, $d$. At $d=5$~nm, BOR-FEM predicts $E_{ex}\approx131$ (black point) together with $Y\approx0.189$ (red point), corresponding to a quantum-yield enhancement of $Y/Y_0\approx9.9$. These values compare favourably with the DDA results of $E_{ex}=130$ and $Y/Y_0\approx9.0$ reported by Khatua \textit{et al.} \cite{Khatua2014}. The BOR-FEM calculations additionally recover the physically expected asymptotic behaviour as the emitter is moved away from the nanorod, with $E_{ex}\rightarrow1$ and $Y\rightarrow Y_0$ for large emitter--nanorod separations, confirming the correct transition from strong near-field coupling to the isolated-emitter limit.
\begin{figure}
	\centering 
	\includegraphics[width = .23\textwidth]{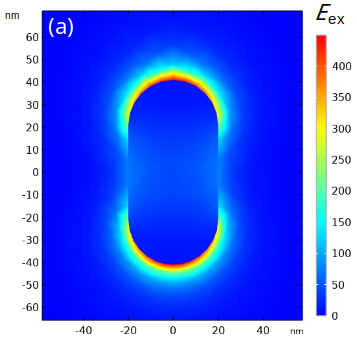}~~~~~~~
	\includegraphics[width = .23\textwidth]{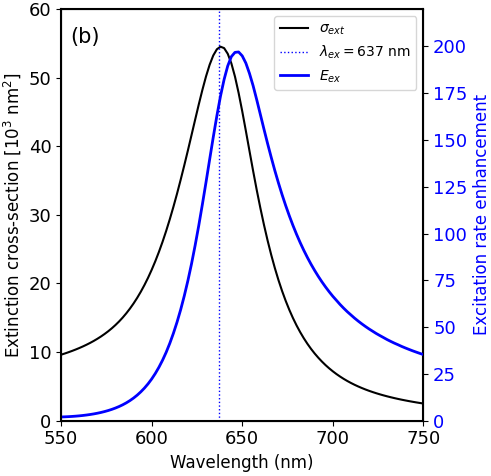} \vspace{0.25cm}\\
	\includegraphics[width = .23\textwidth]{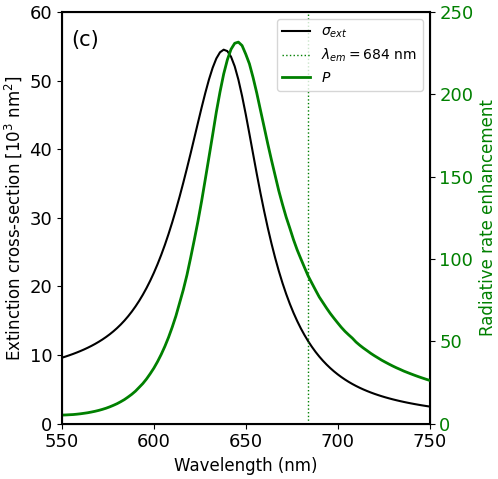}~~~~
	\includegraphics[width = .22\textwidth]{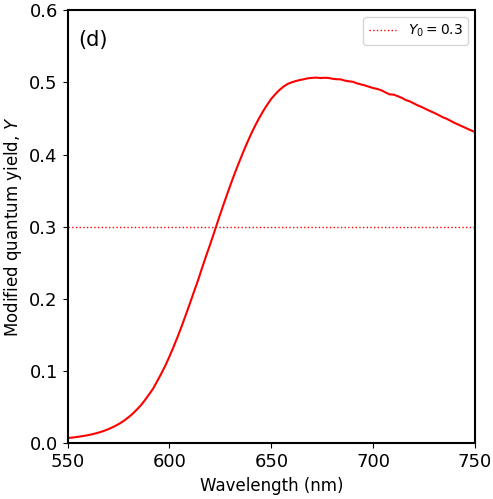}
	\caption{Reproduced results based on BOR-FEM simulations using input data from the Atto655-AuNR system in Ref. \cite{Boom2024}. 
		(a) Near-field map of the excitation rate enhancement evaluated at the excitation wavelength $\lambda_{ex} = 637$ nm. (b) Extinction cross-section of the AuNR (black curve) and wavelength dependence of the excitation rate enhancement at $d = 5$ nm (blue curve). The excitation wavelength is indicated by the dashed vertical line.         
		(c) Extinction cross-section of the AuNR (black curve) and the dependence of the radiative rate enhancement (green curve) on the emission wavelength at $d = 5$ nm. The peak emission wavelength of the emitter is indicated by the dashed vertical line. 
    	(d) Dependence of the modified quantum yield of the emitter on the emission wavelength at $d = 5$ nm. The intrinsic quantum yield $Y_0 = 0.3$ is indicated by the dashed horizontal line.
	}\label{fv3}
\end{figure} 

The normalised radiated and absorbed powers obtained from BOR-FEM at $d=5$ nm are larger than the corresponding values reported from the DDA calculations of Ref.~\cite{Khatua2014}: $P=P_{\mathrm{rad}}/P_0\approx116$ and $P_{\mathrm{abs}}/P_0\approx450$ (BOR-FEM), compared with $P\approx42$ and $P_{\mathrm{abs}}/P_0\approx150$ (DDA). This difference arises because the DDA values were averaged over the entire fluorescence spectrum of the CV molecule, whereas the BOR-FEM results were evaluated at the peak emission wavelength ($\lambda_{\mathrm{em}}=640$ nm), where the plasmonic radiative-rate enhancement is greatest. The more meaningful comparison is therefore through the derived quantities that govern fluorescence enhancement. Although the absolute radiated and absorbed powers differ, both methods predict very similar antenna efficiencies (Table~\ref{t2}), indicating that the ratio of radiative to total emitted power is nearly identical. Consequently, the modified quantum yield and the overall fluorescence enhancement are also in close agreement, despite the differences in the individual power values.

The overall fluorescence enhancement predicted by BOR-FEM is shown in Fig.~\ref{fv}(d). At $d=5$~nm, the model predicts a fluorescence enhancement of $1.3\times10^3$, in close agreement with the value reported by Khatua \textit{et al.} Moreover, the complete distance dependence of the fluorescence enhancement closely follows the published DDA results, demonstrating that BOR-FEM accurately reproduces the combined effects of excitation enhancement, Purcell enhancement, and quantum-yield modification.

\subsubsection{Benchmark II: Atto655 dye molecule near a gold nanorod}
As a second validation, we consider a recent work by Nooteboom et al. \cite{Boom2024}, in which the PEF parameters of a single Atto655 dye molecule coupled to the tip of a cylindrical gold nanorod were calculated using BEM. Unlike the first benchmark, this comparison provides validation against a surface-integral electromagnetic solver that has recently been applied to quantitative PEF calculations.

The benchmark system consists of a cylindrical AuNR with dimensions $L\times W\times W$, where $L=82$~nm and $W=40$~nm (aspect ratio $L/W=2.05$), immersed in water ($n=1.33$). The emitter is represented by a single dipole positioned $5$~nm from the nanorod tip. Excitation occurs at $\lambda_{ex}=637$~nm, while the Atto655 dye emits at $\lambda_{em}=684$~nm. The intrinsic quantum yield and radiative decay rate of Atto655 are $Y_0\approx0.30$ and $\gamma_r^0\approx0.17$~ns$^{-1}$, respectively, corresponding to a free-space radiated power of $P^0_{rad} = \gamma_r^0 hc/\lambda_{em} \approx 49.371$~pW. In the BOR-FEM implementation, a magnetic current source of $I_m\approx1.5327$~V was required to reproduce this radiated power for a dipole located at $r_0=1.0$~nm from the symmetry axis in the absence of the AuNR.

Figure~\ref{fv3}(a) shows the excitation-rate enhancement map calculated using BOR-FEM. The spatial distribution agrees closely with that reported by Nooteboom \textit{et al.} (their Fig.~S4), reproducing both the strong hotspot localisation at the nanorod tips and the rapid spatial decay of the plasmonic near field away from the metal surface. The extinction spectrum obtained from BOR-FEM is shown in Fig.~\ref{fv3}(b), together with the wavelength dependence of the excitation-rate enhancement, while Fig.~\ref{fv3}(c) shows the corresponding radiative rate enhancements compared with the extinction cross-section. The simulated extinction spectrum predicts an LLSPR at approximately $638$~nm, in reasonable agreement with the BEM value of approximately $650$~nm reported in Ref.~\cite{Boom2024}.

The values for $E_{ex}$ and $P$ obtained from BOR-FEM are summarised in Table \ref{t2}. At the emitter position, BOR-FEM predicts $E_{ex}\approx 168$, which is significantly larger than the value reported from the BEM calculations. Since both methods produce very similar near-field intensity distributions, the discrepancy is unlikely to arise from the underlying electromagnetic solution itself. It is likely due to a difference in the evaluation procedure: the published BEM value may have been extracted at a different position on the nanorod surface, such as near the nanorod diameter (blue regions in Fig.~\ref{fv3}(a), where $E_{ex}$ decreases to approximately $50$), rather than along the nanorod axis at the specified emitter position, where the local field reaches its maximum. In contrast, the BOR-FEM value was evaluated directly along the nanorod axis at the specified emitter position, where the local field reaches its maximum.

The Purcell factor obtained from BOR-FEM is also significantly larger than the reported BEM value. This difference is expected because Ref.~\cite{Boom2024} calculated the radiative-rate enhancement by averaging over the fluorescence emission spectrum of Atto655, whereas the BOR-FEM value corresponds to the peak emission wavelength ($\lambda_{em}=684$~nm), where plasmonic enhancement is maximal. The modified quantum yield obtained from BOR-FEM is $Y\approx0.50$, corresponding to a quantum-yield enhancement of $Y/Y_0\approx1.7$, in excellent agreement with the value reported by Nooteboom \textit{et al.} (Table~\ref{t2}). Combining the excitation enhancement and modified quantum yield yields an overall fluorescence enhancement of $F\approx286$. The larger fluorescence enhancement predicted by BOR-FEM therefore arises primarily from the greater excitation-rate enhancement discussed above, rather than from differences in the emission model.
\begin{table}
	\centering
	\resizebox{0.48\textwidth}{!}{%
		\begin{tabular}{|c|c|c|c|c|c|c|c|}
			\hline 
			Method & System & LLSPR (nm)   & $E_{ex}$   & $P$       &  $\eta$  & $Y/Y_0$    & $F$ \\ \hline 
			DDA    & CV-AuNR& $629$   & $130$ & $42^{*}$ &  $22\%$         & $9.0$ & $1170$ \\ 
			BOR-FEM & CV-AuNR& $626$  & $131$ & $116$&  $21\%$         & $9.9$ & $1297$ \\ \hline 
            BEM & Atto655-AuNR& $650$    & $50^{**}$ & $25^{*}$&  $47\%$        & $1.5$ & $75$ \\ 
			BOR-FEM & Atto655-AuNR& $638$ & $168$  & $90$&  $51\%$   & $1.7$ & $286$ \\ \hline 
			FDTD & LH2-AuNR& $900$    & $100$ & $160$&  $57\%$        & $5.5$ & $550$ \\ 
			BOR-FEM & LH2-AuNR& $850$ & $85$  & $212$&  $77\%$   & $7.5$ & $638$ \\ \hline 
		\end{tabular}
	}
	\caption{Summary of the PEF parameters obtained from the model validation. Results of the BOR-FEM simulations are compared to those obtained from DDA, BEM, and FDTD methods at $d = 5$ nm, as reported in Refs. \cite{Khatua2014}, \cite{Wientjes2014}, and \cite{Boom2024}, respectively. $^{*}$These values are spectrally-averaged values over the emitter's emission spectrum in Refs. \cite{Khatua2014,Boom2024}. $^{**}$It appears that the value reported here in Ref. \cite{Boom2024} for $E_{ex}$ is a value near the diameter of the AuNR where the excitation enhancement has dropped considerably compared to its value near the ends of the AuNR. }
	\label{t2} 
\end{table}

Overall, the agreement in the near-field distributions, resonance wavelength, quantum-yield enhancement, and the qualitative trends in the excitation and fluorescence enhancements demonstrates that BOR-FEM reproduces the principal features of the BEM calculations. Together with the DDA benchmark presented above, this provides further confidence in the accuracy and general applicability of the BOR-FEM framework for quantitative PEF simulations.

\subsubsection{Benchmark III: LH2 complex near a gold nanorod}
As a final test of the proposed BOR-FEM method, we considered the light-harvesting complex LH2 from purple bacteria coupled to AuNR, which was investigated experimentally and numerically by Wientjes \textit{et al.}~\cite{Wientjes2014}. Unlike the CV benchmark, this system represents a biologically relevant light-harvesting complex and therefore provides a more stringent test of the BOR-FEM framework under conditions similar to those considered later for LHCII. In Ref.~\cite{Wientjes2014}, the PEF parameters of a single LH2 complex positioned near the tip of AuNR were modelled using FDTD simulations and shown to reproduce the experimentally measured fluorescence enhancement. Here, we examine whether BOR-FEM can reproduce the same physical behaviour despite employing a reduced axisymmetric geometry.
\begin{figure}
	\centering 
	\includegraphics[width = .23\textwidth]{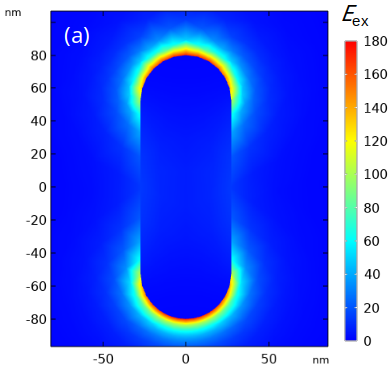}~~~~~~~
	\includegraphics[width = .235\textwidth]{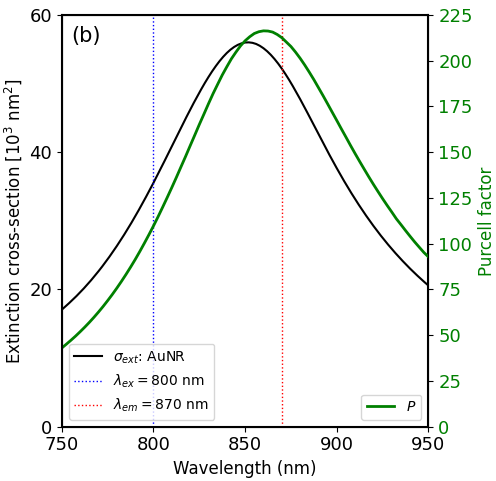} \vspace{0.25cm}\\
	\includegraphics[width = .23\textwidth]{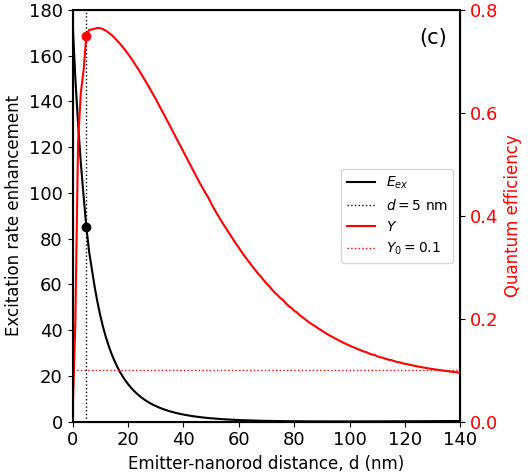}~~~~
	\includegraphics[width = .21\textwidth]{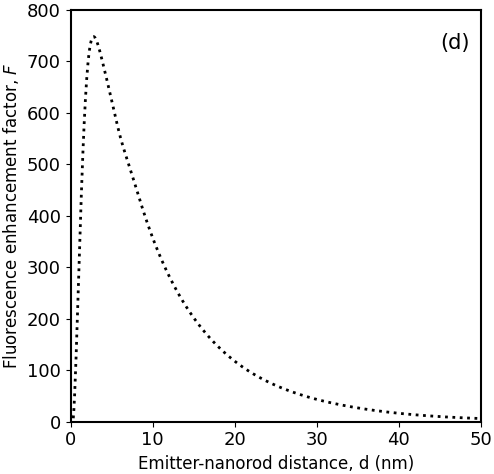}
	\caption{Reproduced results based on BOR-FEM simulations using input data from the LH2-AuNR system in Ref. \cite{Wientjes2014}. 
		(a) Near-field map of the excitation rate enhancement evaluated at the excitation wavelength $\lambda_{ex} = 800$ nm. (b) Extinction cross-section of the AuNR (black curve) and wavelength dependence of the Purcell factor at $d = 5$ nm (green curve). Excitation and emission wavelengths are shown as blue and red dotted lines, respectively.        
		(c) Dependence of the excitation rate enhancement (black curve) at $\lambda_{ex} = 800$ nm and the modified quantum yield $Y$ (red curve) at the emission wavelength $\lambda_{em} = 870$ nm on the emitter-nanorod separation, $d$. The intrinsic quantum yield $Y_0 = 0.1$ is indicated by the dashed horizontal line.
    	(d) Dependence of the fluorescence enhancement factor on the emitter-nanorod distance, evaluated at the excitation and emission wavelengths.
	}\label{fv2}
\end{figure} 

The AuNR used by Wientjes \textit{et al.} has ellipsoidal dimensions of $L\times W\times H$, with $L=160$~nm, $W=60$~nm, and $H=50$~nm. Because this geometry is not rotationally symmetric, it cannot be represented directly within the BOR-FEM formulation. We therefore approximate the nanorod by an equivalent cylindrical geometry with dimensions $L\times W'\times W'$, where the effective diameter is taken as the average of the two transverse axes, i.e., $W' = 0.5(W+H) = 55$ nm, giving an aspect ratio of $L/W' \approx 2.91$. Although this approximation slightly modifies the plasmon resonance, it preserves the overall aspect ratio and enables a meaningful comparison between BOR-FEM and the published FDTD results.

The LH2 complex is excited at $\lambda_{ex}=800$~nm and emits at $\lambda_{em}=870$~nm. Following Ref.~\cite{Wientjes2014}, the intrinsic quantum yield and radiative decay rate of the isolated LH2 complex were taken as $Y_0\approx0.10$ and $\gamma_r^0\approx0.10$~ns$^{-1}$, respectively. These parameters correspond to $P_{rad}^0 = \gamma_{r}^0 hc/\lambda_{em} \approx 22.833$~pW. As in the previous two benchmark studies, the magnetic current source was calibrated to reproduce the radiated power in the absence of the nanorod. For an emitter located at $r_0=1.0$~nm from the symmetry axis, a magnetic current of $I_m\approx1.6543$~V produced the required value of $P_{rad}^0$, thereby establishing the correspondence between the BOR-FEM source and the physical electric dipole representing the LH2 complex.

The excitation-stage validation is shown in Figs.~\ref{fv2}(a,b). Fig.~\ref{fv2}(a) presents the reconstructed near-field map of the excitation-rate enhancement at $\lambda_{ex}=800$~nm. The strongest enhancement is localised near the nanorod surface, where $E_{ex}$ reaches approximately $160$--$180$. At $5$ nm from the AuNR surface, $E_{ex}$ decreases to approximately $80$--$100$, due to a decrease in the local field experienced by the emitter. This behaviour agrees well with the FDTD calculations reported by Wientjes \textit{et al.}, demonstrating that the BOR-FEM model reproduces the expected near-field enhancement profile.

The corresponding spectral response is shown in Fig.~\ref{fv2}(b). The calculated extinction spectrum (black) yields an LLSPR at around $850$ nm, compared with $900$~nm reported from the FDTD simulations. This spectral shift is expected because the BOR-FEM model replaces the original ellipsoidal nanorod by an equivalent cylindrical geometry. Nevertheless, the resonance remains sufficiently close to the excitation and emission wavelengths to preserve the strong spectral overlap required for efficient PEF. The simulated Purcell factor reaches $P\approx212$ at $d=5$~nm, which is likewise comparable to the published FDTD value (Table~\ref{t2}).

The quantitative comparison of the excitation and emission parameters is presented in Figs.~\ref{fv2}(c,d). At the nanorod surface ($d = 0$ nm), $E_{ex}>160$, while at $d=5$~nm BOR-FEM predicts $E_{ex}\approx85$ (black point). The modified quantum yield at this separation is $Y\approx0.75$, corresponding to a quantum-yield enhancement of $Y/Y_0\approx7.5$. Combining these quantities yields a fluorescence enhancement factor of approximately $F\approx638$ (Fig.~\ref{fv2}(d)). Although the simplified cylindrical geometry leads to some quantitative differences in the individual PEF parameters, the predicted excitation enhancement, Purcell enhancement, quantum-yield enhancement, and overall fluorescence enhancement all remain in good agreement with the FDTD simulations and experimental observations reported by Wientjes \textit{et al.} (Table \ref{t2}).

This last benchmark demonstrates that BOR-FEM remains predictive even when the original 3D geometry must be approximated by an equivalent rotationally symmetric structure. The results therefore indicate that the method captures the dominant physical mechanisms governing PEF and can be applied with confidence to rotationally symmetric nanoantenna systems, where full 3D modelling would be computationally more demanding.

\subsection{Application to the LHCII–core–shell AuNR system}
Having validated the BOR-FEM framework against three independent benchmark systems, we now apply it to the experimentally investigated LHCII–core–shell AuNR system reported by Kyeyune \textit{et al.} \cite{Kyeyune2019}. Unlike the benchmark systems, this biohybrid nanoantenna comprises multilayer dielectric architectures, enabling systematic investigation of the influence of shell composition and thickness on PEF. 

Here, the emitter was placed at $r_0 = 0.5$~nm. To determine the value of $I_m$, we calculated $P_{rad}^0$ by approximating the terminal emitter of LHCII as a point dipole using the expression $P_{rad}^0 = \gamma^0_r hc/\lambda_{em} \approx 21.62$~pW, where $\gamma^0_r = 0.074$~ns$^{-1}$, and $\lambda_{em} = 680$ nm \cite{Kyeyune2019}, allowing us to determine the power radiated by the emitter in the absence of the nanorod. Following the calibration procedure described earlier, we obtained $I_m = 4.0225$ V, which produces the same radiated power as the above for the LHCII terminal emitter in the absence of the nanorod. The medium surrounding the emitter-nanorod system was assumed to be water, with a refractive index of $n \approx 1.33$. 
We first consider the spectral overlap between LHCII and the plasmon resonances of the different nanoantenna geometries before analysing the excitation and emission enhancement mechanisms.

\subsubsection{Spectral overlap}
\begin{figure}
	\centering 
	\includegraphics[width = .35\textwidth]{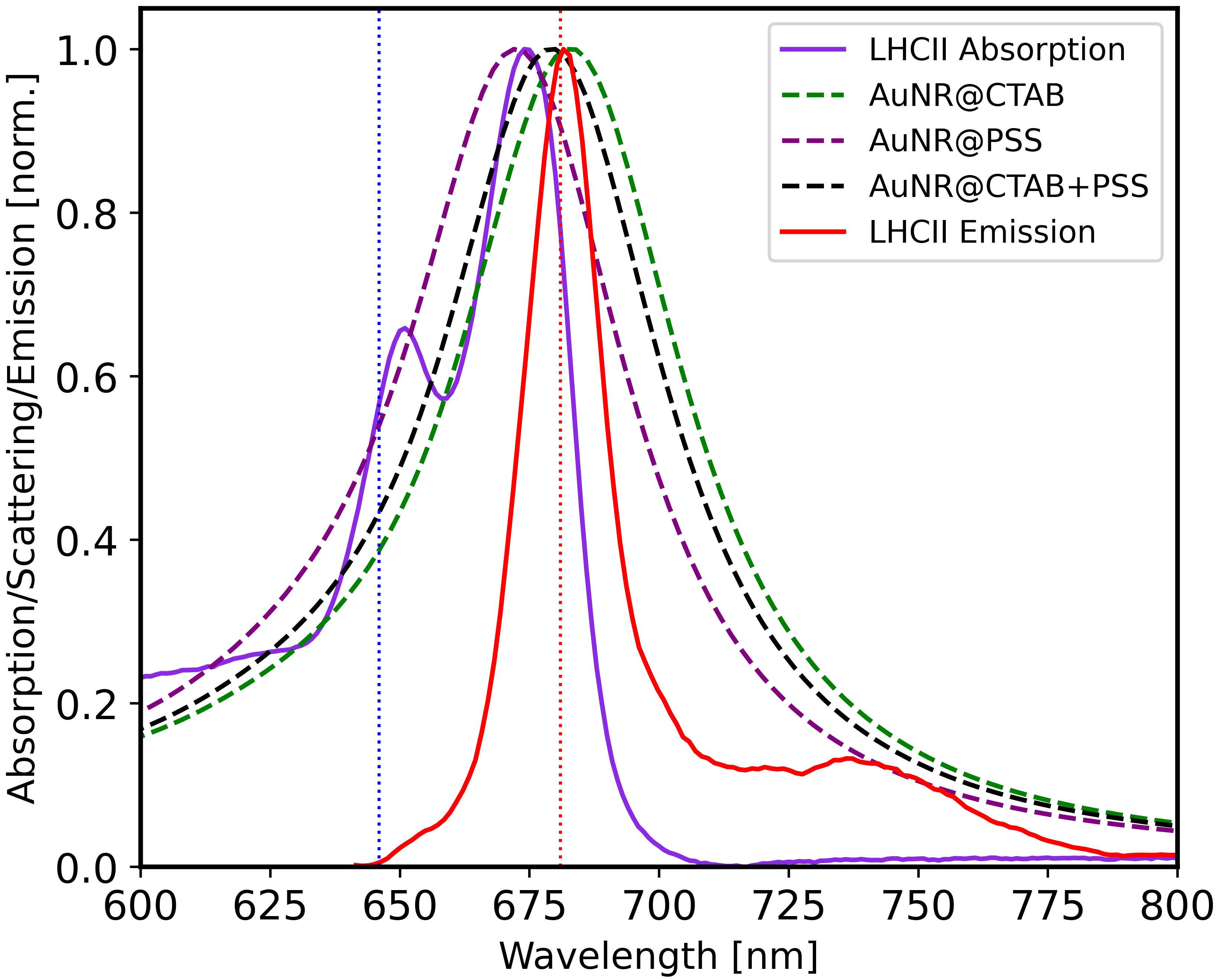}
	\caption{Overlap between LHCII absorption (Q$_{y}$ bands only), LHCII emission, and the scattering spectra of the core-shell nanorods: AuNR@CTAB ($t = 5$ nm), AuNR@PSS ($t = 5$ nm), AuNR@CTAB + PSS ($t_{CTAB} = 3$ nm, $t_{PSS} = 2$ nm). Experimental data for LHCII absorption and emission spectra were taken from Ref. \cite{Kyeyune2019}. 
		The blue dashed vertical line indicates the excitation wavelength $\lambda_{ex} = 646$ nm and the red dashed vertical line indicates the peak emission wavelength  $\lambda_{ex} = 680$ nm of the emitter. }\label{f3}
\end{figure} 

The structural and intrinsic photophysical parameters were taken from the experimental study of Kyeyune \textit{et al.}~\cite{Kyeyune2019}. The intrinsic fluorescence quantum yield of the terminal LHCII emitter was set to $Y_0=0.26$, while the gold nanorod core dimensions were fixed at $L=88$ nm and $W=38$ nm (corresponding to an aspect ratio $L/W\approx2.32$). To systematically map the physical transitions between the near-field quenching, intermediate enhancement, and asymptotic decoupled regimes, the shell thickness of the AuNR with a single dielectric shell was varied continuously from $t = 1$ nm to $t = 300$ nm. To directly evaluate multi-layered architectures, we also modelled a two-layer (CTAB+PSS) dielectric shell, with fixed layer thicknesses of $t_{CTAB} = 3$ nm and $t_{PSS} = 2$ nm, consistent with the configuration in Ref.~\cite{Kyeyune2019}. The complex, wavelength-dependent refractive index of gold was taken from Ref. ~\cite{Christy1972}, while the refractive indices of PSS and CTAB were taken from Refs. \cite{Velas2023} and \cite{Foti2024}, respectively, where we have assumed that the dielectrics have negligible extinction coefficients in the visible region, so that only the real parts of their refractive indices could be used: $n_{PSS} = 1.34$ and $n_{CTAB} = 1.44$. 

The scattering cross-section in Eq.~\eqref{e4} allows us to determine the overlap between the absorption/emission spectra of LHCII and the LLSPR of the core-shell AuNR. 
Fig.~\ref{f3} shows that the Chl \textit{a} Q$_{y}$ transition at $\sim 675$ nm and the Chl \textit{b} Q$_{y}$ transition at $\sim 650$ nm overlap with the broad LLSPR bands of each of the gold core-dielectric shell NRs, with that of AuNR@PSS peaking at $\sim 672$ nm, AuNR@CTAB+PSS at $\sim 679$ nm, and AuNR@CTAB at $\sim 683$ nm. 
The excitation wavelength used experimentally (646 nm) lies close to the peak of the Chl \textit{b} Q$_{y}$ absorption band and strongly overlaps the nanorod plasmon resonance. Consequently, strong excitation enhancement is expected owing to resonant coupling between the excitation wavelength and the nanorod plasmon mode. Excitonic coupling within LHCII subsequently transfers the absorbed energy to the terminal Chl sites responsible for fluorescence emission.

On the other hand, Fig.~\ref{f3} shows that the Chl \textit{a} emission band peaking at $\lambda_{em} \approx 680$ nm (indicated by the red dashed line) more strongly overlaps with the LLSPRs of the AuNR@CTAB and AuNR@CTAB+PSS core-shell NRs compared to the LLSPR of the AuNR@PSS at $\sim 672$ nm. This is caused by differences in the refractive indices of the dielectric shells, which lead to different amounts of dielectric-induced LLSPR shifts. As shown below, the spatial near-field enhancements during both the excitation and emission stages are highly sensitive to these spectral overlap regions. 

To simulate the PEF of the hybrid system, the LHCII complex was assumed to be adsorbed directly onto the outer dielectric shell of the nanorod, as shown in Fig.~\ref{f2}, replicating the experimental configuration reported by Ref.~\cite{Kyeyune2019}. 

\subsubsection{Excitation-rate enhancement}
\begin{figure}
	\centering 
	\includegraphics[width = .23\textwidth]{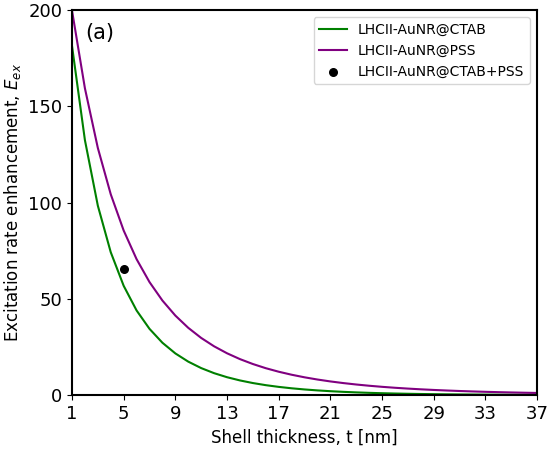}~~~~
	\includegraphics[width = .23\textwidth]{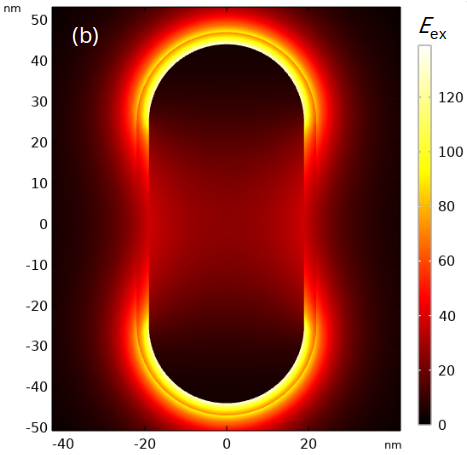}
	\caption{(a) Dependence of the excitation-rate enhancement of the Q$_{y}$ bands of LHCII  on the shell thickness, $t$, of the core-shell nanorods: AuNR@CTAB ($t$ is varied), AuNR@PSS ($t$ is varied), AuNR@CTAB + PSS ($t$ is fixed with $t_{CTAB} = 3$ nm, $t_{PSS} = 2$ nm), obtained using Eq. ~\eqref{e5}. 
		(b) Near-field map of the excitation-rate enhancement of the single LHCII molecule when it interacts with a two-layer dielectric shell-gold core nanorod: AuNR@CTAB + PSS ($t_{CTAB} = 3$ nm, $t_{PSS} = 2$ nm) at the excitation wavelength $\lambda_{ex} = 646$ nm.
	}\label{f4}
\end{figure} 

The excitation-rate enhancement, $E_{ex}$, experienced by the Chl \textit{b} Q$_{y}$ transition in LHCII in the vicinity of a core--shell nanorod is shown in Fig.~\ref{f4}(a) as a function of shell thickness for the single-layer geometries (AuNR@CTAB and AuNR@PSS), and compared with the dual-shell configuration (AuNR@CTAB+PSS), for which the shell thicknesses are fixed at the experimentally studied values of $t_{CTAB} = 3$ nm and $t_{PSS} = 2$ nm reported in Ref.~\cite{Kyeyune2019}. Excitation enhancement is largest in the thin-shell regime, where the emitter remains strongly coupled to the plasmon-enhanced near field of the nanorod. Specifically, at a shell thickness of $t = 1$ nm, $E_{ex}$ reaches $\sim 180$ and $\sim 200$ for AuNR@CTAB and AuNR@PSS, respectively. As the shell thickness increases, $E_{ex}$ decreases monotonically owing to the spatial attenuation of the localised plasmonic field. The larger excitation enhancement observed for AuNR@PSS relative to AuNR@CTAB arises from the stronger spectral overlap between its LLSPR peak ($\sim 672$~nm) and the LHCII excitation wavelength (see Fig.~\ref{f3}), resulting in a larger local electric field at the emitter position.

The dual-layer (AuNR@CTAB+PSS) geometry yields an excitation-rate enhancement of $E_{ex} \approx 65$ at a thickness of $t = 5$ nm, which is in close agreement with the experimentally inferred value of $E_{ex} \approx 63$ reported by Ref.~\cite{Kyeyune2019}. In contrast, the corresponding single-shell geometries predict somewhat smaller or larger enhancement factors ($E_{ex} \approx 57$ for AuNR@CTAB and $E_{ex} \approx 85$ for AuNR@PSS). This result highlights the importance of incorporating the experimentally relevant multilayer shell architecture when modelling fluorescence enhancement in this system.

The reconstructed 3D excitation-enhancement distribution for the AuNR@CTAB+PSS geometry is shown in  Fig.~\ref{f4}(b). The field map is obtained directly from the 2D BOR-FEM solution through rotational reconstruction about the symmetry axis. Pronounced electromagnetic hotspots are localised at the metal–dielectric and dielectric–dielectric interfaces, reflecting the strong confinement of the plasmonic near field. The largest excitation-rate enhancements occur near the AuNR–CTAB interface, where $E_{ex} > 120$, reflecting the strong confinement of the plasmonic near field at the metal–dielectric boundary. Significant enhancement is also maintained throughout the dielectric-shell region, where $E_{ex} \sim 80-120$. At the outer PSS–water interface, where the LHCII complex resides experimentally, the excitation-rate enhancement remains substantial ($E_{ex} \sim 40-80$) before decaying into the surrounding medium. These results demonstrate that the fluorescence enhancement observed in the LHCII–nanorod system is driven primarily by near-field coupling to localised plasmonic modes, while also showing that appreciable excitation enhancement extends well beyond the immediate vicinity of the metal surface.

\subsubsection{Radiative rate enhancement}
\begin{figure}
	\centering 
	\includegraphics[width = .25\textwidth]{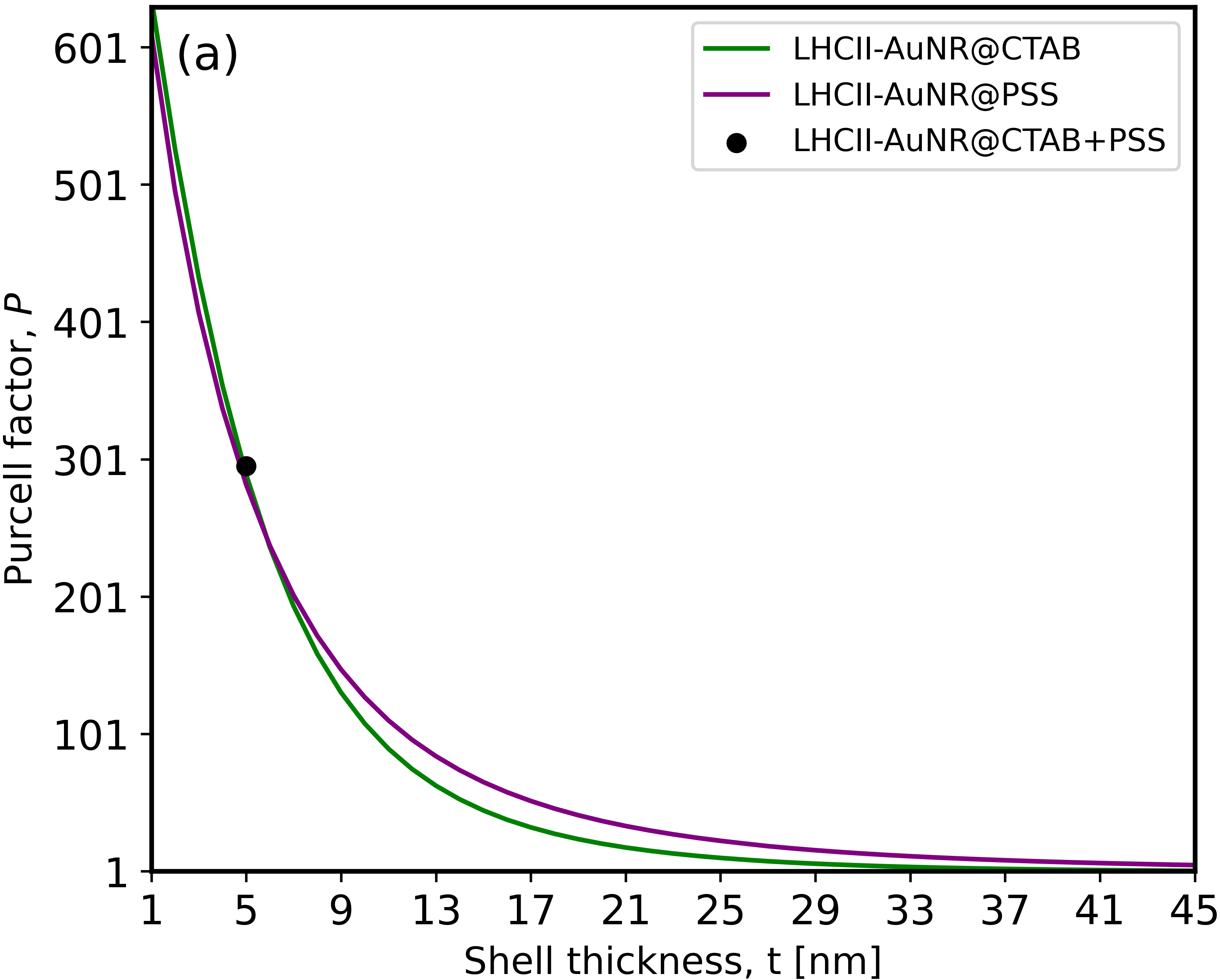}~~~~
	\includegraphics[width = .22\textwidth]{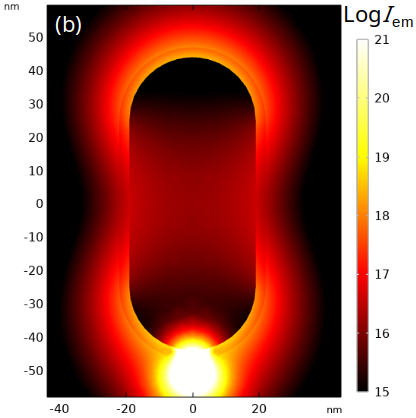}
	\caption{(a) Dependence of the radiative decay rate enhancement of the LHCII terminal emitter on the shell thickness, $t$, of the core-shell nanorods: AuNR@CTAB ($t$ is varied), AuNR@PSS ($t$ is varied), AuNR@CTAB + PSS ($t$ is fixed with $t_{CTAB} = 3$ nm, $t_{PSS} = 2$ nm).
		(b) Enlarged near-field map of the intensity distribution at the emitter's position (brightest region) and around the two-layer dielectric shell-gold core nanorod: AuNR@CTAB + PSS ($t_{CTAB} = 3$ nm, $t_{PSS} = 2$ nm) at the emission wavelength $\lambda_{em} = 680$ nm.
	}\label{f5}
\end{figure} 

Fig.~\ref{f5}(a) shows the dependence of the radiative decay rate enhancement (Purcell factor, $P$) of the emitting LHCII dipole on the shell thickness, $t$, in the emission stage. The Purcell factor depends on the power radiated by the emitting dipole in the presence of a nanoantenna \cite{Moha2008,Khatua2014}. In the presence of the core-shell NR, the dipole radiated power is enhanced due to the scattered field of the nanoshell, which is dominated by the induced dipole field of the AuNR. As the nanoshell thickness increases, the scattered field reaching the emitter decreases due to dielectric screening, thereby reducing the Purcell factor. Consequently, as shown in Fig.~\ref{f5}(a), $P$ decreases monotonically with increasing shell thickness and approaches unity in the thick-shell limit, where the emitter behaves as if it were isolated from the nanoantenna. 
The large Purcell factors at small shell thickness in Fig.~\ref{f5}(a) show that the local density of optical states experienced by the LHCII dipole is substantially enhanced. 
In contrast to the excitation enhancement, the Purcell factors associated with the different shell compositions are very similar because all structures exhibit LLSPRs that are nearly on-resonant with the peak emission wavelength of LHCII. Thus, the Purcell enhancement is governed primarily by emitter–nanoantenna separation rather than by small differences in resonance wavelength. At $t \approx5$ nm, all geometries yield  $P  \approx 296$, which as a similar order of magnitude as the experimentally inferred value of $P  \approx 200$~\cite{Kyeyune2019}.  

Fig.~\ref{f5}(b) shows the near-field emission-intensity distribution for the AuNR@CTAB+PSS geometry, where the intensity is calculated as $I_{em} = \frac{1}{2}nc\epsilon_{0}|\mathbf{E}|^2$, where $\epsilon_0$ is the permittivity of free space and $|\mathbf{E}|$ is the absolute value of the total electric field at the emission wavelength. The emitter's position is on the outer PSS shell, consistent with the experimental configuration. The emitted field of the dipole interferes constructively with the scattered field of the core-shell NR, creating a localised plasmonic hotspot (brightest region in Fig.~\ref{f5}(b)) at the emitter's position. Strong field confinement is also observed near the AuNR–CTAB interface, where the electromagnetic intensity is substantially higher than at the CTAB–PSS interface owing to the short decay length of localised surface plasmons. These results further confirm that radiative-rate enhancement originates from near-field coupling between the emitter and the plasmonic nanoantenna. 

\subsubsection{Antenna efficiency and modified quantum yield}
\begin{figure}
	\centering 
	\includegraphics[width = .45\textwidth]{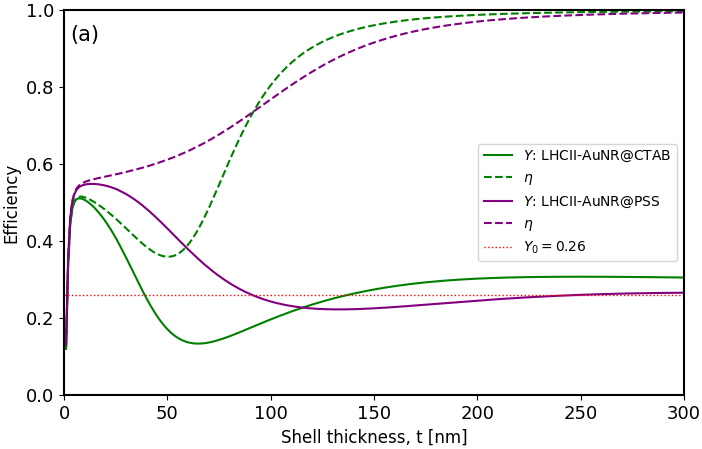}\vspace{0.4cm}\\
	\includegraphics[width = .45\textwidth]{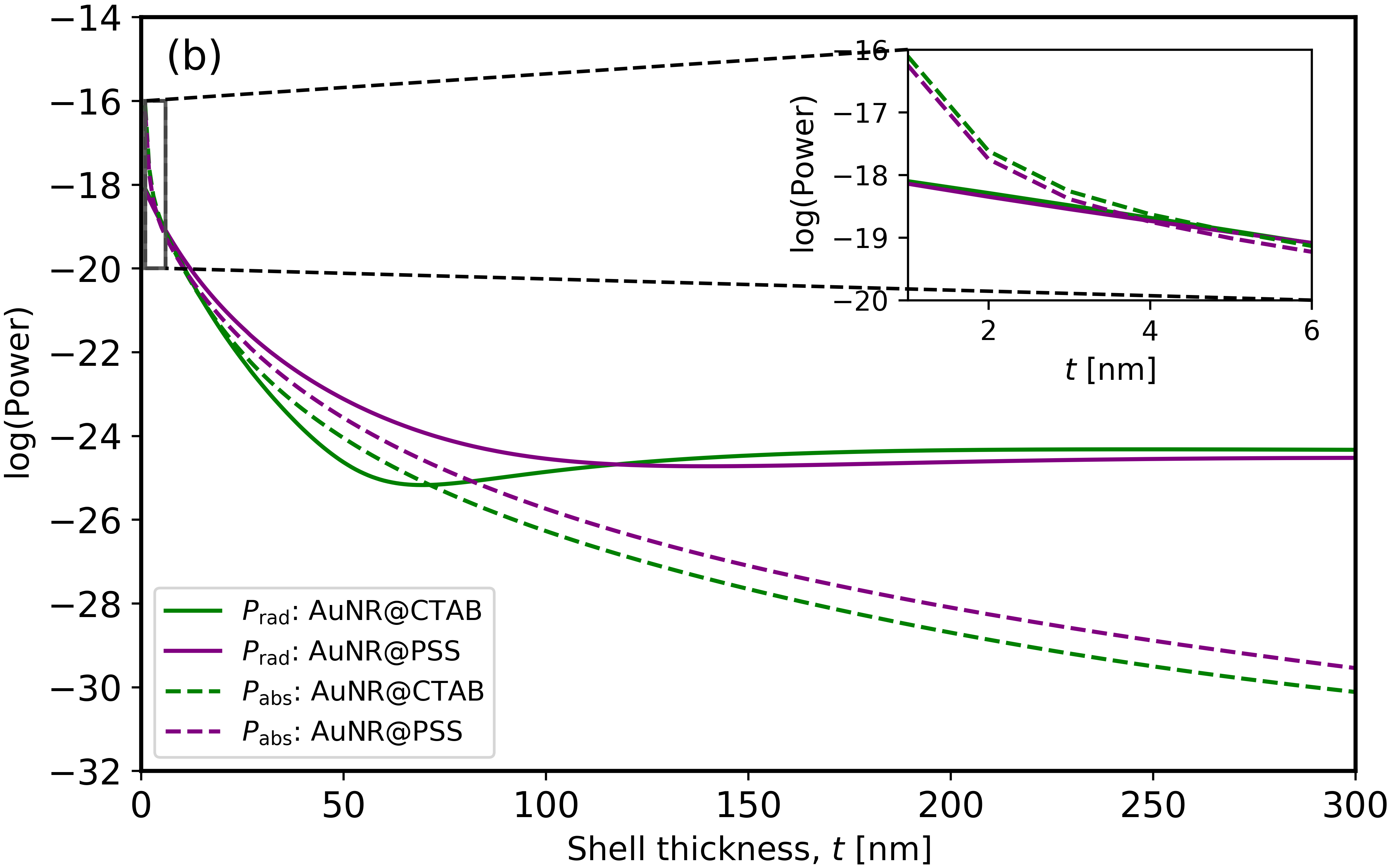}
	\caption{(a) Dependence of the antenna efficiencies of the nanoshells and the modified quantum yields of the LHCII dipole emitter on the shell thickness, $t$, of the core-shell nanorods: AuNR@CTAB and AuNR@PSS.
		(b) Dependence of the radiated and absorbed powers due to the two core-shell nanorods on the shell thickness. Since $log(Power)$ follows the same trend as the power, it is used here to show hidden behaviour at small and large $t$ values. 
	}\label{f6}
\end{figure}

Fig.~\ref{f6}(a) shows the dependence of the antenna efficiency $\eta$ and modified quantum yield $Y$ on shell thickness for the different core–shell nanorods. Four distinct regimes can be identified.

In the extremely thin-shell limit ($t\le 1.5$ nm), the emitter is strongly coupled to the metallic core and non-radiative energy transfer dominates. As shown by the absorbed powers in the inset of Fig.~\ref{f6}(b), $P_{abs}$ is very high in this regime, resulting in $\eta < Y_0$ and consequently $Y <Y_0$. Irrespective of the shell dielectric, the system therefore operates in the \emph{emission quenching} regime. 

As the shell thickness increases beyond $\sim 1.5$ nm, the system enters a thin-shell \emph{emission enhancement} regime. In this region, both $\eta$ of the nanoshells and $Y$ of the emitter increase rapidly with increasing shell thickness and reach their maximum values at $t\sim 6$ nm. In this \emph{thin-shell limit}, the corresponding core-volume fraction is $f_{c} = 1/(1 + (2t/L))(1+(2t/W))^{2} \ge 0.5$ and the emitter remains strongly coupled to the plasmonic near field. At the same time, absorptive losses decrease rapidly with increasing separation from the metal surface. Because the Purcell factor $P$ remains very large throughout this regime (Fig.~\ref{f5}(a)), $(1-Y_0)/P \rightarrow 0$ in Eq. \eqref{e3}, leading to $Y \approx \eta$. Under these conditions, the emitter-nanoshell system behaves as an efficient two-way transmitter-receiver antenna \cite{Novot2007}, because the absorbed power $P_{abs}$ decreases more rapidly with increasing shell thickness than the radiated power $P_{rad}$, as can be seen in the steepness of the curves in the inset of Fig.~\ref{f6}(b).

Beyond the thin-shell limit ($f_c<0.5$), the Purcell factor decreases sufficiently that the intrinsic emitter quantum yield can no longer be neglected because Eq. \eqref{e3} has a non-vanishing contribution from $(1-Y_0)/P$. Consequently, $\eta$ and $Y$ begin to diverge. Fig.~\ref{f6}(a) shows that for AuNR@CTAB, both quantities decrease with increasing $t$, with $\eta$ showing a dip near $t = 50$ nm, beyond which $\eta \rightarrow 1$. 
This dip represents the physical transition from a near-field coupling regime, dominated by Ohmic losses in the gold core, to a dielectric screening regime, in which the thick shell alters out-coupling and radiation damping. 
As shown in Fig.~\ref{f6}(b), for AuNR@CTAB, $P_{rad}$ decreases more rapidly than $P_{abs}$ in this regime, causing a reduction in $\eta$ and a corresponding decrease in $Y$ to below $Y_0$. The system therefore operates in the \emph{emission quenching} regime.

Beyond $t \sim 50$ nm, the nanoshell becomes very thick ($f_c << 0.5$), and dielectric screening progressively isolates the emitter from the gold core. In this limit, $P_{abs}$ decreases more rapidly than $P_{rad}$, causing $\eta$ and $Y$ to increase once again (Fig.~\ref{f6}(a)). As $t$ continues to increase, absorptive losses vanish ($P_{abs}\rightarrow 0$) due to complete dielectric screening of the AuNR, while the radiated power approaches that of the isolated emitter ($P_{rad}\rightarrow P_{rad}^{0}$). As a result, $\eta \rightarrow 1$ and $P \rightarrow 1$, and the modified quantum yield approaches its intrinsic value ($Y \rightarrow Y_0$). 

The same qualitative behaviour is observed for AuNR@PSS, although the dip is less pronounced and shifted to a larger shell thickness ($t \approx 100$ nm). This difference arises because the refractive index of PSS ($n = 1.34$) is much closer to that of the surrounding medium ($n=1.33$) than that of CTAB, leading to a much weaker dielectric screening. Hence, $Y \rightarrow  Y_0$ more rapidly with increasing shell thickness compared to the values of $Y$ obtained for AuNR@CTAB (Fig.~\ref{f6}(a)). Notably, this trend is consistent with our earlier prediction for $Y$ \cite{Ugwuoke2021} based on the Gersten-Nitzan model \cite{Nitzan1981}, which is found here to best describe the simulated behaviour of the LHCII–AuNR@PSS system.

\subsubsection{Fluorescence enhancement}
\begin{figure}
	\centering 
	\includegraphics[width = .27\textwidth]{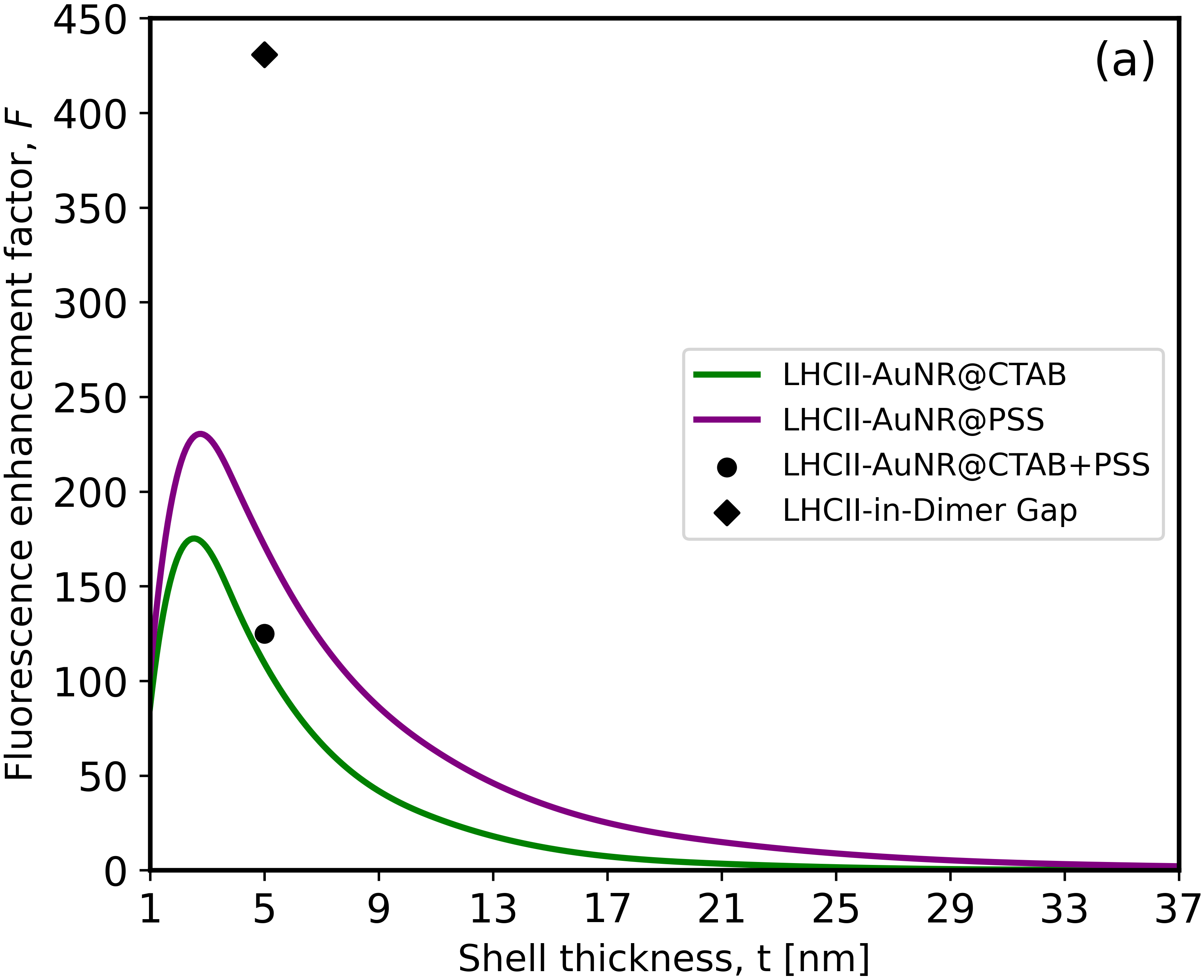}~~
	\includegraphics[width = .22\textwidth]{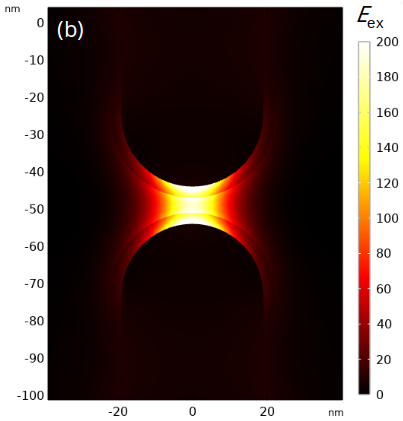}
	\caption{(a) Dependence of the fluorescence enhancement factor, $F$,  of the LHCII dipole emitter on the shell thickness, $t$, of the core-shell nanorods: AuNR@CTAB and AuNR@PSS. The values of $F$ obtained from the LHCII-AuNR@CTAB+PSS and the AuNR@CTAB+PSS-LHCII-AuNR@CTAB+PSS systems (LHCII-in-Dimer Gap) at $t_{CTAB} = 3$ nm and $t_{PSS} = 2$ nm are also indicated. 
		(b) Near-field map of the excitation rate-enhancement obtained with the LHCII dipole emitter in the gap of a AuNR@CTAB+PSS dimer. 
	}\label{f7}
\end{figure}

The fluorescence enhancement factor, $F$, combines the effects of excitation-rate enhancement and quantum-yield modification and therefore provides the most direct measure of the overall performance of a plasmonic nanoantenna. 
Fig.~\ref{f7}(a) shows the dependence of $F$ on the shell thickness $t$ for the different core-shell nanorod geometries considered in this work. To further assess the predictive capability of the BOR-FEM framework, we also calculated the excitation and emission enhancement factors of an LHCII dipole emitter in the nanogap of a AuNR@CTAB+PSS core-double shell dimer. The near-field map of the excitation-rate enhancement due to the nanogap is shown in Fig.~\ref{f7}(b). The enhancement factors are compared with the experimental results of Ref. \cite{Kyeyune2019} and summarised in Table \ref{t1}. 

For both single-shell architectures, the fluorescence enhancement exhibits a pronounced maximum at intermediate shell thicknesses. The AuNR@CTAB geometry reaches an optimal enhancement factor of  $F\approx 177$ at $t\approx 2.5$ nm, whereas AuNR@PSS reaches an optimal value of $F\approx 232$ at $t\approx 3$ nm. 
The superior performance of the PSS shell originates primarily from its larger excitation-rate enhancement (Fig.~\ref{f4}), which results from the closer spectral overlap between its LLSPR and the excitation wavelength of LHCII. The existence of an optimum shell thickness reflects the competing distance dependences of excitation enhancement and non-radiative quenching. At very small separations, excitation enhancement is large but strong energy transfer to the metal suppresses the modified quantum yield. At large separations, quenching is reduced, but the plasmon-enhanced near field becomes too weak to sustain significant excitation enhancement. Maximum fluorescence enhancement therefore occurs at an intermediate shell thickness where these competing mechanisms are optimally balanced. 

Although the AuNR@CTAB+PSS structure yields a lower fluorescence enhancement ($F\approx130$) than the optimum values obtained for the single-shell geometries, it provides the most physically realistic representation of the experimentally studied nanoantenna. Most importantly, the calculated excitation-rate enhancement $(E_{ex}\approx65$) agrees remarkably well with the experimentally inferred value ($E_{ex}\approx63$). This result highlights the importance of explicitly incorporating the multilayer dielectric environment surrounding the nanorod, rather than approximating the system using a single effective dielectric shell.
\begin{table}
	\centering
	\resizebox{0.48\textwidth}{!}{%
		\begin{tabular}{|c|c|c|c|c|c|}
			\hline
			Enhancement & \multicolumn{4}{c|}{Simulations} & \multicolumn{1}{c|}{Experiment} \\ \cline{2-6}
			factors at& AuNR@CTAB & AuNR@PSS & AuNR@CTAB+PSS & Nanoshell dimer & AuNR@CTAB+PSS  \\ 
			$\lambda_{ex} = 646$ nm, & $t = 2.5$ nm & $t = 3.0$ nm & $t_{CTAB} = 3.0$ nm & $t_{CTAB} = 3.0$ nm & $t_{CTAB} = 2.9$ nm \\
			$\lambda_{em} = 680$ nm	&    &  & $t_{PSS} = 2.0$ nm &  $t_{PSS} = 2.0$ nm  & $t_{PSS} = 1.9$ nm \\
			\hline
			$E_{ex}(\lambda_{ex})$ & 200 & 127 & 65 & 200 & $\sim$ 63 \\ \hline
			$P(\lambda_{em})$ & 450  & 400  & 296 & 440  & $\sim$ 200 \\
			$Y(\lambda_{em})/Y_0$ & 1.54  & 1.83 & 2.00  & 2.15  & $\sim$ 3.80 \\ \hline
			$F(\lambda_{ex},\lambda_{em})$ & 177 & 232 & 130 & 430  & $\sim$ 242 \\
			\hline
		\end{tabular}
	}
	\caption{Simulated values of the enhancement factors for LHCII-AuNR@Dielectric shell with the dielectrics CTAB, PSS, and CTAB+PSS, and for LHCII-in-Dimer Gap (AuNR@CTAB+PSS nanoshell dimer), in comparison with experimental values reported in Ref. \cite{Kyeyune2019}. Here, the value of $E_{ex}$ near the hotspot (see Fig.~\ref{f7}(b)) has been used for the dimer gap nanoantenna. }
	\label{t1} 
\end{table}
The reconstructed excitation-enhancement map for the dimer geometry (Fig.~\ref{f7}(b)) reveals the formation of a strongly confined plasmonic hotspot within the nanogap. 
Excitation-rate enhancements vary from $E_{ex}\sim60$ in the outer regions of the gap to values exceeding $E_{ex}\sim200$ near the hotspot center. When combined with the corresponding emission enhancement factors, these local field enhancements yield fluorescence enhancement factors approaching $F\sim 430$ for an emitter near the nanogap hotspot. 
This is comparable to twice the value inferred previously from experimental data ($F\sim 242$). 

In contrast, the AuNR@CTAB+PSS model reproduces the experimentally inferred value for $E_{ex}$ with considerably greater fidelity, although it underestimates both $Y$ and $F$. This discrepancy can be traced largely to the treatment of non-radiative losses. In the experimental analysis of Ref.~\cite{Kyeyune2019}, absorptive losses within the nanoantenna were assumed to be negligible, leading to an inferred quantum yield of $Y\approx0.98$. The BOR-FEM calculations explicitly account for Ohmic dissipation within the gold nanorod and therefore predict substantially lower quantum yields in the thin-shell regime ($1.5$ nm $\le t \le 6$ nm). The results suggest that non-radiative plasmonic losses remain an important component of the fluorescence-modification process and should not be neglected when quantitatively interpreting fluorescence enhancement experiments. Overall, the close agreement obtained for the experimentally relevant multilayer geometry demonstrates that BOR-FEM provides a physically realistic and computationally efficient framework for predicting fluorescence enhancement in rotationally symmetric emitter–nanoantenna systems.

\section{Conclusion}
We have employed and validated a BOR-FEM framework for modelling PEF in emitter–nanoantenna systems. By exploiting rotational symmetry, or more generally an equivalent axisymmetric representation of the electromagnetic problem, the method substantially reduces the computational cost of conventional 3D electromagnetic simulations while preserving the key optical processes governing excitation enhancement, Purcell enhancement, antenna efficiency, quantum-yield modification, and overall fluorescence enhancement. Validation against three previously reported emitter–-NR systems demonstrated excellent agreement with published DDA, BEM, and FDTD simulations, establishing BOR-FEM as a reliable framework for quantitative PEF modelling.

Application of the method to LHCII coupled to gold core–-dielectric-shell nanorods revealed that fluorescence enhancement is governed by the interplay between excitation enhancement and emission modification, both of which depend strongly on shell composition and shell thickness. Four distinct regimes were identified: an extreme thin-shell quenching regime dominated by non-radiative losses, a thin-shell enhancement regime characterised by strong near-field coupling and high antenna efficiency, an intermediate thick-shell regime in which dielectric screening suppresses emission enhancement, and an asymptotic decoupling regime where the intrinsic quantum yield of the emitter is recovered. The experimentally relevant AuNR@CTAB+PSS architecture reproduced the excitation enhancement reported experimentally and yielded fluorescence enhancement factors consistent with measured values. Comparisons between single-shell, dual-shell, and dimer configurations further demonstrated the strong sensitivity of PEF to dielectric stratification and emitter placement.

Beyond the systems investigated here, the BOR-FEM framework is applicable to a broad class of plasmonic and dielectric nanoantenna problems for which an equivalent axisymmetric transformation can be constructed. Because the method can be implemented using standard finite-element software and readily accommodates multilayer nanoantenna architectures without requiring the emitter dipole moment as a direct input parameter, it provides an efficient and accessible platform for the quantitative design and optimisation of PEF in molecular, biological, and nanoscale photonic systems.




\section*{Acknowledgements}
This work acknowledges funding from the South African Department of Science, Technology, and Innovation (DSTI) through the South African Quantum Technology Initiative (SA QuTI), Stellenbosch University (SU), and the National Research Foundation (NRF).

\bibliographystyle{unsrtnat} 
\bibliography{references}
\end{document}